\documentclass[final,3p,times]{elsarticle}

\usepackage{lipsum}
\usepackage{mathrsfs}
\usepackage{amsmath}
\usepackage{amssymb}
\usepackage{braket}
\usepackage{graphicx}
\usepackage{xcolor}
\usepackage{bm}
\usepackage{comment}

\journal{Nuclear Physics B}

\begin{document}

\begin{frontmatter}



\title{Residual gauge symmetry and color confinement \\ in the Yang-Mills theory}


\author[first]{Naoki Fukushima}
\affiliation[first]{organization={Department of Physics, Graduate School of Science and Engineering, Chiba University},
    postcode={263-8522},
      state={Chiba},
      country={Japan}}
\author[second1,second2]{Kei-Ichi Kondo}
\affiliation[second1]{organization={Department of Physics, Graduate School of Science, Chiba University},
      postcode={263-8522},
      state={Chiba},
      country={Japan}}
\affiliation[second2]{organization={Research and Education Center for Natural Sciences, Keio University},
      postcode={223-8521},
      state={Kanagawa},
      country={Japan}}

\begin{abstract}
We examine the restoration of the residual gauge symmetry in the Yang-Mills theory to be regarded as a confinement criterion.
For this purpose we restrict the four-dimensional $SU(2)$ Yang-Mills instantons to those with spatial spherical symmetry $SO(3)$, which automatically causes the dimensional reduction of the four-dimensional $SU(2)$ Yang-Mills theory to the two-dimensional $U(1)$ gauge-scalar theory as implemented explicitly by the Witten transformation.
In this setting, we show that the restoration of the residual gauge symmetry  occurs due to condensations of instantons and anti-instantons, although the residual gauge symmetry was spontaneously broken in the perturbative vacuum.
This result demonstrates that the true confinement phase is a disordered phase in which all internal electric symmetries of the gauge field remain unbroken.


\end{abstract}




\end{frontmatter}



\allowdisplaybreaks

\section{Introduction}
\label{introduction}

In the previous papers \cite{Kondo-Fukushima,Fukushima-Kondo}, we investigated the restoration of the \textit{residual gauge symmetry} (RGS) in the Yang-Mills theory to give a general criterion for confinement by taking into account topological configurations. The study of RGS in light of color confinement was initiated by Hata \cite{Hata} who reproduced the Kugo-Ojima color confinement criterion in the Lorenz gauge \cite{Kugo-Ojima, Kugo-Ojima-prove}.
The RGS is the gauge symmetry remaining even after imposing the gauge fixing condition.
Although the RGS is ``spontaneously broken'' in the perturbative vacuum, the RGS should be restored in the true confining vacuum, since the confinement phase is regarded as a disordered phase in which all internal electric symmetries of the gauge field remain unbroken.

In this paper, we examine the restoration of the RGS in the $SU(2)$ Yang-Mills theory in the $D=4$ Euclidean spacetime due to topological configurations which are expected to contribute also to quark confinement.
We suppose that such topological configurations in the pure Yang-Mills theory are given by a class of instantons with a certain symmetry, from which the important topological configurations such as magnetic monopoles or vortices are derived.
It is known that instantons as solutions of the Yang-Mills  self-dual equation can respect various spacetime and spatial symmetries.
However, the ordinary instantons with the \textit{four-dimensional spacetime spherical symmetry}  in $D = 4$  Euclidean spacetime are generally considered to be irrelevant for confinement.

In view of these, the instantons we consider in this paper to be relevant to RGS and confinement are those with the \textit{three-dimensional spatial spherical symmetry} $SO(3)$, which corresponds to the cylindrical symmetry (or axial symmetry around the Euclidean time axis) from the perspective of $D = 4$ spacetime.
This restriction automatically causes the dimensional reduction of the $D = 4$ $SU(2)$ Yang-Mills theory to the $D = 2$ $U(1)$ gauge-scalar theory with a quartic scalar self-interaction.
This fact was first found in the study of finding multi-instanton solutions by Witten \cite{Witten1977} and later proved by Forgacs and Manton \cite{Forgacs-Manton} (see also section 4.3 of \cite{Manton}).
In the dimensionally reduced $D = 2$ $U(1)$ gauge-scalar theory, it is proved that vortex solutions, i.e., $D=2$ instanton solutions exist, see section 7.14.3 of \cite{Manton}.

It turns out that adopted symmetric instantons not only makes the $D=4$ Euclidean Yang-Mills action (integral) finite, but also gives a nontrivial minimal action carrying the minimal topological charge and therefore
they can be considered one of the most effective and important configurations for the path integral.
In fact, we show that the global $U(1)$ symmetry associated with the RGS as a subgroup of the original $SU(2)$ symmetry is restored by taking into account the non-perturbative effects due to condensations of instantons and anti-instantons in the dimensionally reduced $D = 2$ $U(1)$ gauge-scalar theory, although the RGS is spontaneously broken in the perturbative vacuum.
This result confirms that the true confinement phase is a disordered phase in which all internal electric symmetries of the gauge field  remain unbroken.
This result should be compared with the other confinement criteria: the Wilson criterion for quark confinement \cite{Wilson74} and the Kugo-Ojima criterion for color confinement.



\section{Gauge fixing and the residual gauge symmetry}

\subsection{$SU (2)$ Yang-Mills theory and gauge fixing}

We start from the $SU(2)$ Yang-Mills theory in the four-dimensional ($D = 4$) Euclidean space specified by the following action supplemented by a topological term with an angle $\vartheta \in \mathbb{R}$:
\begin{align}
 S_{\text{YM}} = \int d^4 x \ \mathscr{L}_{\text{YM}} , \ \mathscr{L}_{\text{YM}} = \frac{1}{4} \mathscr{F}_{\mu \nu}^A \mathscr{F}_{\mu \nu}^A + \frac{i \vartheta}{32 \pi^2} \mathscr{F}_{\mu \nu}^A {}^* \mathscr{F}_{\mu \nu}^A,
\end{align}
where the Yang-Mills gauge field $\mathscr{A}_\mu(x)$ is defined as the Lie algebra valued field $\mathscr{A}_\mu(x)=\mathscr{A}_\mu^A(x)T_A$ with the generators $T_A = \frac{1}{2} \sigma_A$ of the Lie algebra $su(2)$ defined by the Pauli matrices $\sigma_A$ ($A=1,2,3$) for the $SU(2)$ gauge group,
$\mathscr{F}_{\mu \nu}$ is the Lie algebra $su(2)$ valued field strength of the gauge field $\mathscr{A}_\mu$ defined by
$\mathscr{F}_{\mu \nu} := \partial_\mu \mathscr{A}_\nu - \partial_\nu \mathscr{A}_\mu - i [\mathscr{A}_\mu , \mathscr{A}_\nu] = - \mathscr{F}_{\nu \mu}$ and ${}^* \mathscr{F}_{\mu \nu}$ represents the Hodge dual of $\mathscr{F}_{\mu \nu}$, i.e.,
${}^* \mathscr{F}_{\mu \nu}:=\frac12 \varepsilon_{\mu \nu \rho \sigma} \mathscr{F}_{\rho \sigma}$.

We introduce the \textit{generalized} gauge transformation by the Lie-group element
$ 
U (x) = e^{i g \omega (x)}, \ \omega (x) = \omega^A (x) T_A
$ 
of the gauge group $G$ with the Lie-algebra valued transformation function $\omega (x) = \omega^A (x) T_A$: For the gauge field $\mathscr{A}_\mu (x)$ and the matter field $\varphi (x)$,
 \begin{align}
  \delta_U \mathscr{A}_\mu (x) &:=  U(x) (\mathscr{A}_\mu(x) + i g^{- 1} \partial_\mu) U^\dagger(x) - \mathscr{A}_\mu(x) ,
   \nonumber\\
  \delta_U \varphi (x) &:= U (x) \varphi (x) - \varphi(x) .
  \label{eq:gauge-transformation-A,phi}
 \end{align}
  Similarly, for the other fields: the Nakanishi-Lautrup auxiliary field $\mathscr{B} (x)$, the Faddeev-Popov ghost field $\mathscr{C} (x)$ and antighost field $\bar{\mathscr{C}} (x)$, we define the generalized gauge transformation:
 \begin{align}
  \delta_U \mathscr{B} (x) &:= U (x) \mathscr{B}(x) U^\dagger (x) - \mathscr{B}(x), \nonumber\\
  \delta_U \mathscr{C} (x) &:= U (x) \mathscr{C}(x) U^\dagger (x) - \mathscr{C}(x), \nonumber\\
  \delta_U \bar{\mathscr{C}} (x) &:= U (x) \bar{\mathscr{C}}(x) U^\dagger (x) - \bar{\mathscr{C}}(x).
  \label{eq:gauge-transformation-B,C}
 \end{align}
The infinitesimal version of the generalized local gauge transformation is given by
\begin{align}
  \delta_\omega \mathscr{A}_\mu (x)
&= \partial_\mu \omega (x) + g \mathscr{A}_\mu (x) \times \omega (x)
:= \mathscr{D}_\mu \omega (x),
\quad  \delta_\omega \varphi (x)
= i g \omega (x) \varphi (x) , \nonumber\\
  \delta_\omega \mathscr{B} (x)
&= g \mathscr{B}(x) \times \omega(x), 
\quad   \delta_\omega \mathscr{C} (x)
  = g \mathscr{C}(x) \times \omega(x), 
\quad
\delta_\omega \bar{\mathscr{C}} (x)
   = g \bar{\mathscr{C}} (x)\times \omega(x) ,
\label{eq:infinitesimal_gt}
\end{align}
In what follows we use the notation $(\mathscr{A}\times \mathscr{B})^A = f^{ABC} \mathscr{A}^B \mathscr{B}^C$ with the structure constant $f^{A B C} = \varepsilon^{A B C}$ for $S U (2)$ group.

As a gauge-fixing (GF) condition, we adopt the Lorenz gauge:
\begin{align}
\partial^\mu \mathscr{A}_\mu(x) = 0.
\end{align}
Then the total Lagrangian density $\mathscr{L}$ consisting of the gauge-invariant term $\mathscr{L}_{\text{inv}}$ and the GF plus the Faddeev-Popov (FP) ghost term $\mathscr{L}_{\text{GF+FP}}$  is given by
\begin{align}
  \mathscr{L} = \mathscr{L}_{\text{inv}} + \mathscr{L}_{\text{GF+FP}} , \quad 
  \mathscr{L}_{\text{inv}} = \mathscr{L}_{\text{YM}}
   + \mathscr{L}_{\text{matter}} (\varphi , D_\mu \varphi) ,
  \ \mathscr{L}_{\text{GF+FP}} = - i \delta_{\text{B}} \left[\text{tr} \left\{\bar{\mathscr{C}} (\partial^\mu \mathscr{A}_\mu + \frac{\alpha}{2} \mathscr{B})\right\}\right],
\end{align}
where $D_\mu$ is the covariant derivative of the matter field $\varphi$ in the fundamental representation: $D_\mu \varphi := \partial_\mu \varphi - i g \mathscr{A}_\mu \varphi$ and $\delta_{\text{B}}$ is the Becchi-Rouet-Stora-Tyutin (BRST) transformation.
We say that the RGS exists for a gauge fixing condition if the gauge transformed field $\mathscr{A}_\mu^U(x):=\Omega_\mu(x) + U(x) \mathscr{A}_\mu(x) U^\dagger(x)$ of $\mathscr{A}_\mu(x)$ also satisfies the same gauge fixing condition
where we defined the pure gauge form $\Omega_\mu (x)$ as
\begin{align}
 \Omega_\mu (x) := i U (x) \partial_\mu U^\dagger (x) .
\end{align}
In the Lorenz gauge $\partial^\mu \mathscr{A}_\mu(x) = 0$, therefore, the RGS exists if a non-trivial solution $U \in SU (2)$ exists for the equation:
\begin{align}
 \partial^\mu (\Omega_\mu(x) + U(x) \mathscr{A}_\mu(x) U^\dagger(x)) = 0, 
 \label{eq:residual_symmetry1}
\end{align}
for a given gauge field $\mathscr{A}_\mu(x)$ satisfying the Lorenz gauge $\partial^\mu \mathscr{A}_\mu(x) = 0$.

Remember that any element $U$ of the gauge group $SU (2)$ can be expressed as
\begin{align}
 U (x) &= \exp \left(i \omega^A (x) \frac{\sigma_A}{2}\right) = \exp \left(i \theta (x) \text{n}^A (x) \frac{\sigma_A}{2}\right) = \cos \frac{\theta (x)}{2} + i \sin \frac{\theta (x)}{2} \text{n}^A (x) \sigma_A \in SU(2) \quad (A = 1 , 2 , 3) ,
 \label{eq:SU(2)}
\end{align}
where $\mathbf{n} (x)=(\text{n}^A (x))_{A=1,2,3}$ is a unit vector field  satisfying $\mathbf{n} (x) \cdot \mathbf{n} (x) := \text{n}^A (x) \text{n}^A (x) = 1$ and $\theta (x)$ is an angle of the local rotation around the axis along $\mathbf{n} (x)$.
Then, the pure gauge form $\Omega_\mu (x)$ is represented as
\begin{align}
 \Omega_\mu (x) = i U (x) \partial_\mu U^\dagger (x) = \frac{\bm{\sigma}}{2} \cdot [\partial_\mu \theta (x) \mathbf{n} (x) + \sin \theta (x) \partial_\mu \mathbf{n} (x) + (1 - \cos \theta (x)) (\partial_\mu \mathbf{n} (x) \times \mathbf{n} (x))] .
\end{align}
Notice that $\mathbf{n} , \ \partial_\mu \mathbf{n}$ and $\partial_\mu \mathbf{n} \times \mathbf{n}$ constitute orthogonal bases of $su (2)$.


\subsection{Symmetry in the quantum gauge theory}

In classical theory, the symmetry is usually understood as the transformation which leaves the action invariant, which is called the symmetry of the action.
In quantum theory, however, the classical field is replaced by the operator acting on the state vector which does not appear in the classical theory.
Therefore, we can extend the concept of the symmetry without restricting to the symmetry of the action by taking into account the state space.
In the manifest Lorentz covariant formalism of the quantum gauge field theory,
we can consider the symmetry as the transformation of the field operator restricted to the physical state space $\mathcal{V}_{\text{phys}}:= \{ |{\rm phys} \rangle; Q_B|{\rm phys} \rangle=0 \}$ as a subspace of the whole state space $\mathcal{V}$ where $Q_B$ is the BRST charge operator. Here, the BRST symmetry is the nilpotent symmetry of the quantum theory: $Q_{\text{B}}^2 \equiv 0$.
Therefore, the concept of the symmetry can be extended to the symmetry conserved only in the physical subspace  that can be applied to any choice of the gauge-fixing condition.

The infinitesimal generalized gauge transformation $\delta_\omega$ of the total Lagrangian density $\mathscr{L}$ by the gauge transformation function $\omega$ can be written in a BRST-exact form \cite{Hata,Fukushima-Kondo}:
\begin{align}
  \delta_\omega \mathscr{L} = \delta_\omega \mathscr{L}_{\text{GF+FP}} = i \delta_{\text{B}} F , \quad F= (\mathscr{D}_\mu \bar{\mathscr{C}})^A  \partial^\mu \omega^A .
\end{align}
This is because
\begin{align}
  \delta_\omega \mathscr{L} = \delta_\omega \mathscr{L}_{\text{GF+FP}}
  &= - i \delta_\omega \delta_{\text{B}} \left[\bar{\mathscr{C}}^A \left(\partial^\mu \mathscr{A}_\mu^A + \frac{\alpha}{2} \mathscr{B}^A\right)\right] = - i \delta_{\text{B}} \delta_\omega \left[\bar{\mathscr{C}}^A \left(\partial^\mu \mathscr{A}_\mu^A + \frac{\alpha}{2} \mathscr{B}^A\right)\right] = - i \delta_{\text{B}} \delta_\omega [\bar{\mathscr{C}}^A \partial^\mu \mathscr{A}_\mu^A] \nonumber\\
  &= i \delta_{\text{B}} [(\mathscr{D}_\mu \bar{\mathscr{C}})^A] \partial^\mu \omega^A ,
\end{align}
where we have used the fact that $\delta_{\text{B}}$ and $\delta_\omega$ commute \cite{Kondo-Fukushima}.
By this reason, according to Hata, we do not need to consider the invariance of the gauge fixing and the associated ghost term $\mathscr{L}_{\text{GF+FP}}$ written in the BRST-exact form in the following discussions of the RGS restoration.

\subsection{Residual gauge symmetry in the preceding work}\label{sec:RGS_pr}

In order for the infinitesimal gauge transformation $\delta_\omega$ with the Lie algebra valued gauge transformation function $\omega(x) = \omega^A(x) T_A$ ($T_A$ is the generator of the Lie algebra) to become the RGS in the Lorenz gauge, $\omega^A (x)$ must satisfy the equation:
\begin{align}
  \partial^\mu \delta_\omega \mathscr{A}_\mu^A(x) = 0 \Rightarrow \ \partial^\mu \mathscr{D}_\mu^{AB} \omega^B (x) = 0,
  \label{eq:RGS_noncompact}
\end{align}
where $\mathscr{D}_\mu^{AB}$ is the covariant derivative in the adjoint representation: $\mathscr{D}_\mu^{AB} = \delta^{AB} \partial_\mu + \varepsilon^{ACB} \mathscr{A}_\mu^C$.
A solution of this equation \eqref{eq:RGS_noncompact}  is given by\cite{Hata}
\begin{align}
  \omega^A (x) = \delta^{A E} b_\mu x^\mu + c^A,
  \label{eq:omega_noncompact}
\end{align}
with a specific index $E$ of the Lie algebra.
Here $x$-independent part $c^A$ corresponds to the color rotation and $x$-independent coefficient $b_\mu$ plays the role of the (infinitesimal) global parameter for  defining the global symmetry associated with the RGS  according to the Noether method.
Then the gauge transformation of the total Lagrangian density $\mathscr{L}$ under the  gauge transformation function $\omega (x)$ with (\ref{eq:omega_noncompact}) is given by the BRST exact form \cite{Hata}:
\begin{align}
  \delta_\omega \mathscr{L}
= i \delta_{\text{B}} [(\mathscr{D}_\mu \bar{\mathscr{C}})^A] \delta^{A E} b^\mu = i \delta_{\text{B}} [(\mathscr{D}_\mu \bar{\mathscr{C}})^E] b^\mu .
\end{align}

\section{Symmetric instanton and dimensional reduction}

It is rather difficult to solve the equation (\ref{eq:residual_symmetry1}) for a finite gauge transformation $U(x)$ to take into account the topological effects beyond the infinitesimal case $\omega(x)$  which does not include the topological configurations.
Therefore, some simplification is necessary to proceed this analysis.

\subsection{Witten transformation (Ansatz)}

For this purpose, we adopt the transformation (change of variables) with the $D=4$ cylindrical symmetry (or $D=3$ spherical symmetry) which we call the Witten transformation:
\begin{align}
 -\mathscr{A}_4^A (x) =& \frac{x_A}{r} A_0 (r , t) , \nonumber\\
 \ -\mathscr{A}_j^A (x) =& \frac{x_j x_A}{r^2} A_1 (r , t) + \frac{\delta_{j A} r^2 - x_j x_A}{r^3} \varphi_1 (r , t) + \frac{\varepsilon_{j A k} x_k}{r^2} [1 + \varphi_2 (r , t)] \ (j = 1 , 2 , 3).
 \label{eq:Witten_Ansatz}
\end{align}
This transformation changes the $D=4$ $SU(2)$ gauge field variables $\mathscr{A}_\mu^A(x)$ $(\mu=1,2,3,4; A=1,2,3)$ into the $D=2$ field variables $A_0(r , t)$, $A_1 (r , t)$, $\varphi_1(r , t)$, $\varphi_2(r , t)$ depending only on the Euclidean time $t$ and the spatial radius $r$ defined by $r := \sqrt{x_1^2 + x_2^2 + x_3^2}$.

This transformation was originally used by Witten \cite{Witten1977} as an Ansatz to solve the self-dual equation to obtain multi-instanton solutions for $D=4$ Yang-Mills theory.
A remarkable advantage of this transformation is to achieve the dimensional reduction as will be explained shortly.

The tensor structures appearing in this transformation follow from a specific choice of a unit vector field $\mathbf{n}=\mathbf{x}/r$:
\begin{align}
  \text{n}^A(x) = \frac{x_A}{r} \Rightarrow \ \partial_0 \text{n}^A = 0 , \ (\partial_0 n \times n)^A = 0 , \ \partial_j \text{n}^A = \frac{\delta_{j A} r^2 - x_j x_A}{r^3} , \ (\partial_j n \times n)^A = \frac{\varepsilon_{j A k} x_k}{r^2} .
  \label{eq:choiceofn}
\end{align}
The inverse of the Witten transformation to write the $D=2$ field in terms of $D = 4$ field is obtained  as
\begin{align}
A_0(r,t)= \frac{x_A}{r} \mathscr{A}_4^A (x) , \quad
A_1(r,t)= \frac{x_j}{r} \frac{x_A}{r} \mathscr{A}_j^A (x) , \quad
\varphi_1(r,t) = \frac{\delta_{j A} r^2 - x_j x_A}{2 r} \mathscr{A}_j^A(x)  , \quad
\varphi_2(r,t)=\varepsilon_{A j k} \frac{x_k}{2} \mathscr{A}_j^A(x) - 1 .
\end{align}

The Witten transformation is the most general form respecting both $SU(2)$ internal symmetry and spatial $SO(3)$ symmetry.
When preserving the $SO(3)$ spatial symmetry within the theory, the Witten Ansatz is the natural choice.
Preserving spatial $SO(3)$ symmetry determines the dependence of the field on the angular variable in polar coordinates.
Thus, the gauge degrees of freedom uniquely determine the Ansatz.
As a consequence of required symmetries, dimensional reduction occurs naturally.
See \cite{Forgacs-Manton} and the section 4.3 of \cite{Manton} for the details.
Moreover, the $D = 4$ Yang-Mills action remains finite even after applying this transformation.

It should be remarked that a simpler example of the dimensional reduction by imposing $x_2$- and $x_3$-independence of gauge field
\begin{align}
  \mathscr{A}_\mu^A = \mathscr{A}_\mu^A (t , x_1)
\end{align}
cannot be used for our purpose
because this reduction leads to the divergent $D = 4$ Yang-Mills action due to the integral over $x_2$ and $x_3$ coordinates and hence such gauge fields do not contribute to the path integral,  although it respects the original full $SU (2)$ gauge symmetry.

\subsection{Dimensional reduction through the Witten transformation}

The field strength $\mathscr{F}_{\mu\nu}^A$ is expressed under the Witten transformation (\ref{eq:Witten_Ansatz}) as
\begin{align}
  - \frac{\varepsilon_{j k \ell}}{2} \mathscr{F}_{k \ell}^A = & \frac{x_j x_A }{r^2} \frac{\varphi_1^2 + \varphi_2^2-1}{r^2} - \frac{\delta_{j A} r^2 - x_j x_A }{r^3} D_1 \varphi_2 + \frac{\varepsilon_{j A k} x_k}{r^2} D_1 \varphi_1 , \nonumber\\
 - \mathscr{F}_{4 j}^A = & \frac{x_j x_A}{r^2} F_{0 1} + \frac{\delta_{j A} r^2 - x_j x_A}{r^3} D_0 \varphi_1 + \frac{\varepsilon_{j A k} x_k}{r^2} D_0 \varphi_2 ,
  \label{F}
\end{align}
where we have defined
\begin{align}
 D_\mu \varphi_a := \partial_\mu \varphi_a + \varepsilon_{a b} A_\mu \varphi_b , \ F_{\mu \nu} := \partial_\mu A_\nu - \partial_\nu A_\mu \ (\mu , \nu = 0 , 1 , \ a , b = 1 , 2) .
\end{align}
In what follows, we use the notation:
\begin{align}
\partial_0 := \frac{\partial}{\partial t}, \quad
\partial_1 := \frac{\partial}{\partial r}.
\end{align}
Then, the $D=4$ Yang-Mills action with the Lagrangian density
\begin{align}
\mathscr{L}_{\text{YM}} = \frac{1}{2} (\mathscr{F}_{4 j}^A)^2 + \frac{1}{2} \left(\frac{1}{2} \varepsilon_{jk \ell} \mathscr{F}_{k \ell}^A\right)^2 + \frac{i \vartheta}{64 \pi^2} \varepsilon_{\mu \nu \rho \sigma} \mathscr{F}_{\mu \nu}^A \mathscr{F}_{\rho \sigma}^A
\end{align}
is rewritten in terms of the reduced functions:
\begin{align}
 S_{\text{YM}}[\mathscr{A}] = &4 \pi \int_{- \infty}^\infty dt \int_{0}^\infty d r \ \mathscr{L}_{\text{GS}} = 4\pi S_{\text{GS}}[A , \psi] := \tilde S_{\text{GS}}[A , \psi],
 \nonumber\\
 \mathscr{L}_{\text{GS}} = &D_\mu \varphi_a D_\mu \varphi_a + \frac{1}{4} r^2 F_{\mu \nu} F_{\mu \nu} + \frac{1}{2 r^2}(\varphi_a \varphi_a - 1)^2 + \frac{i \vartheta}{16 \pi^2} \varepsilon_{\mu \nu} F_{\mu \nu} \nonumber\\
 = &(D_\mu \psi)^* D_\mu \psi + \frac{1}{4} r^2 F_{\mu \nu} F_{\mu \nu} + \frac{1}{2 r^2}(|\psi|^2 - 1)^2 + \frac{i \vartheta}{16 \pi^2} \varepsilon_{\mu \nu} F_{\mu \nu},
 \label{YM-L}
\end{align}
where we have defined the complex-valued scalar field $\psi := \varphi_1 + i \varphi_2$ to obtain $D_\mu \varphi_a D_\mu \varphi_a = (D_\mu \psi)^* D_\mu \psi$ and ignored the total derivative term \cite{Witten1977}.
Thus the $SU (2)$ Yang-Mills theory witten in terms of $\mathscr{A}$ in $D=4$ Euclidean space is reduced to the $U(1)$ gauge-scalar theory written in terms of $A , \psi$ in $D=2$ spacetime of the coordinates $(r,t)$ with a curved metric $g_{\mu\nu}= r^{- 2} \delta_{\mu\nu}$ implying a negative constant curvature.
In this case, the vacuum expectation value of the operator $\mathscr{O} [\mathscr{A}] = \mathscr{O} [A , \psi]$ in the path integral formulation has the correspondence between the original $D=4$ theory and the reduced $D=2$ theory:
\begin{align}
  Z_4^{- 1} \int d \mathscr{A} \ e^{- S_{\text{YM}} [\mathscr{A}]} \mathscr{O} [\mathscr{A}] \rightarrow \ Z_2^{- 1} \int d A d \psi \ e^{- 4\pi S_{\text{GS}} [A , \psi]} \mathscr{O} [A , \psi] ,
\end{align}
with the respective partition function:
\begin{align}
  Z_4 = \int d \mathscr{A} \ e^{- S_{\text{YM}} [\mathscr{A}]} \rightarrow Z_2=\int d A d \psi \ e^{-4\pi S_{\text{GS}} [A , \psi]} .
\end{align}

In order to understand the correspondence between the $D=4$ $SU(2)$ Yang-Mills theory and the associated $D=2$ $U(1)$ gauge-scalar theory furthermore,
we perform a specific $SU(2)$ gauge transformation with the cylindrical symmetry expressed by
\begin{align}
U = \exp \left(i \theta (r , t) \frac{x_A}{r} \frac{\sigma_A}{2}\right) = \cos \frac{\theta (r , t)}{2} + i \sin \frac{\theta (r , t)}{2} \frac{x_A}{r} \sigma_A \in SU(2) .
\label{gt-U1}
\end{align}
Under this transformation (\ref{gt-U1}), the gauge field $\mathscr{A}_\mu$ is transformed as\cite{Witten1977}
\begin{align}
 \mathscr{A}_4 \rightarrow \ \mathscr{A}_4^\prime &=U \mathscr{A}_4 U^\dagger + i U \partial_0 U^\dagger 
 =- \frac{\sigma_A}{2} \frac{x_A}{r} \left(A_0 - \partial_0 \theta\right) , \nonumber\\
  \mathscr{A}_j \rightarrow \ \mathscr{A}_j^\prime &=U \mathscr{A}_j U^\dagger + i U \partial_j U^\dagger \nonumber\\
  &= - \frac{\sigma_A}{2} \Biggl[ \frac{x_j x_A}{r^2} (A_1 - \partial_1 \theta) +  \frac{\delta_{j A} r^2-x_j x_A}{r^3} (\varphi_1 \cos \theta + \varphi_2 \sin \theta) + \frac{\varepsilon_{j A k} x_k}{r^2} (1 - \varphi_1 \sin \theta + \varphi_2 \cos \theta ) \Biggr],
 \label{U1-gt}
\end{align}
where we have used $\frac{x_A}{r} \sigma_A \frac{x_B}{r} \sigma_B = \frac{x_A}{r} \frac{x_B}{r} (\delta_{A B} I + i \varepsilon_{A B C} \sigma_C) = I$ and $\partial_j =\frac{\partial}{\partial x^j} =\frac{\partial r}{\partial x^j} \frac{\partial}{\partial r} =\frac{x_j}{r} \frac{\partial}{\partial r} = \frac{x_j}{r} \partial_1$.
Then, the \textit{finite} gauge transformation by a general element (\ref{gt-U1}) of $SU(2)$ is equivalent to a \textit{finite} $U(1)$ gauge transformation in $D = 2$ Abelian $U(1)$ gauge-scalar theory as explicitly shown by Actor \cite{Actor79}:
\begin{align}
 A_\mu &\rightarrow \ A_\mu^\prime = A_\mu - \partial_\mu \theta \ \Leftrightarrow \ A_\mu \rightarrow \ A_\mu^\prime =e^{- i \theta} (A_\mu + i \partial_\mu) e^{i \theta} , \nonumber\\
 \begin{pmatrix}
  \varphi_1 \\
  \varphi_2
 \end{pmatrix}
 &\rightarrow \
 \begin{pmatrix}
  \varphi_1^\prime \\
  \varphi_2^\prime
 \end{pmatrix}
= \begin{pmatrix}
  \cos \theta & \sin \theta \\
  - \sin \theta & \cos \theta
 \end{pmatrix}
 \begin{pmatrix}
  \varphi_1 \\
  \varphi_2
 \end{pmatrix}
 \ \Leftrightarrow \ \psi \rightarrow \ \psi^\prime = e^{- i \theta} \psi.
 \label{eq:gauge_transformation_Witten_Ansatz}
\end{align}
This observation indicates that the gauge symmetry of the original compact non-Abelian  group $SU(2)$ is reduced to a compact Abelian subgroup $U(1)$ associated with the dimensional reduction.
The reason can be understood from the gauge-covariant decomposition for the gauge field which includes the Witten transformation as a special case, see \cite{PR}.

\section{Dimensional reduction of the residual gauge symmetry}

\subsection{\label{sec:GF_RGS}gauge fixing and the residual gauge symmetry from new perspective}

In what follows, we focus on the vacuum configuration to discuss the spontaneous breaking of the RGS  in the perturbative vacuum and its restoration in the true vacuum with topological effects being included.
Starting at the perturbative vacuum $\mathscr{A}_\mu(x) = 0$ which corresponds to $A_0 = A_1 = \varphi_1 = 0$ and $\varphi_2 = - 1$, the equation \eqref{eq:residual_symmetry1} for $U \in SU (2)$ with the parameterization (\ref{gt-U1}) reduces to the equation for a single field variable $\theta(r,t)$ after the dimensional reduction \cite{Fukushima-Kondo}:
\begin{align}
 \partial_\mu (r^2 \partial_\mu \theta (r,t)) - 2 \sin \theta (r,t) = 0 \ (\mu=0,1).
 \label{eq:gfc-perturbative1}
\end{align}
However, this is a partial differential equation with respect to two variables $r$ and $t$, which is too difficult to be solved for the whole range.
Therefore we assume the existence of a solution to equation \eqref{eq:gfc-perturbative1}, because it is not the main issue to obtain the solution explicitly,
Rather, we observe that the solution of the second-order differential equation \eqref{eq:gfc-perturbative1} must have global parameters coming from the choice of boundary conditions, which is a common feature in solving the differential equation irrespective of the linear or non-linear differential equation.
These global parameters play the role of the (infinitesimal) global parameter in defining the global $U (1)$ symmetry according to the Noether method, which is to be broken or not to be broken, see \cite{Fukushima-Kondo} for more details.


This observation can be explicitly checked in the $t$-independent case where the above equation reduces to the Gribov equation:
\begin{align}
 \partial_1 (r^2 \partial_1 \theta (r)) - 2 \sin \theta (r) = 0.
 \label{eq:gfc-perturbative}
\end{align}
It is helpful to rewrite this equation to see the global behavior of the solution.
By introducing a new variable (a virtual time) $\tau = \ln r$, \eqref{eq:gfc-perturbative} is rewritten into
\begin{align}
 \ddot{\theta} (\tau) + \dot{\theta} (\tau) - 2 \sin \theta (\tau) = 0 ,
\end{align}
where the dot denotes the differentiation with respect to $\tau$.
This is identical to the equation of motion for the damped pendulum of Gribov \cite{Gribov}.
Therefore, it is obvious that the solution for $\theta (r)$ indeed exists for any $\tau$ and hence for any $r$, as indeed shown by Gribov \cite{Gribov}.
Thus, the gauge transformation $U$ satisfying \eqref{eq:gfc-perturbative} exists.

In order to see the behavior of the solution of the equation \eqref{eq:gfc-perturbative}, we linearize the equation around $\theta = \theta_0$ by substituting $\theta (r) = \theta_0 + \delta (r)$ and by retaining the linear term alone in $\delta (r)$, \eqref{eq:gfc-perturbative} reduces to the second order linear differential equation of the Euler type:
\begin{align}
 r^2 \delta^{\prime \prime}(r) + 2 r \delta^{\prime}(r) - 2 \cos \theta_0 \delta(r) = 2 \sin \theta_0,
\end{align}
where the prime denotes the differentiation with respect to $r$. Therefore, it is easy to see that the asymptotic solution for $r \rightarrow \ 0$ and $r \rightarrow \ \infty$ is given by
\begin{align}
 \theta (r) \sim
 \begin{cases}
  \theta_0 - \tan \theta_0 + C_1 r^{\frac{1}{2}(- 1 + \sqrt{1 + 8 \cos \theta_0})} &
  (1 + 8 \cos \theta_0 > 0 , \ r \rightarrow \ 0) \\
  \theta_0 - \tan \theta_0 + C_2 r^{- \frac{1}{2}} \cos \left(\frac{1}{2}\sqrt{|1 + 8 \cos \theta_0|} \ln r + \alpha\right) &
   (1 + 8 \cos \theta_0 < 0 , \ r \rightarrow \ \infty)
 \end{cases}
 .
\end{align}
By further imposing a boundary condition, e.g., $U \to 1$, i.e. $\theta \rightarrow \ 0 \ (r \rightarrow \ 0)$, and $\theta \rightarrow \ \pm \pi \ (r \rightarrow \infty)$ based on the physical considerations of the usual damped pendulum, we obtain the linearized solution:
\begin{align}
 \theta (r) \sim
 \begin{cases}
 C_1 r \ &(r \rightarrow \ 0) \\
  \ \pm \pi + C_2 r^{- \frac{1}{2}} \cos \left(\frac{1}{2}\sqrt{7} \ln r + \alpha\right) \ &(r \rightarrow \ \infty)
 \end{cases} .
  \label{eq:solution_Gribov}
\end{align}
The solution $\theta$ obtained in the finite case corresponds to $\omega$ of \eqref{eq:omega_noncompact} obtained in the infinitesimal case and hence the coefficients $C_1$ or $C_2$ corresponds to the global parameter $b_\mu$ in \eqref{eq:omega_noncompact}.
This analysis shows that the solution has global parameters, e.g., $C_1$ and $C_2$ coming from the choice of boundary conditions needed to determine the solution of the differential equation.

\begin{figure}[t]
\centering
    \centering
    \includegraphics[width=0.7\columnwidth]{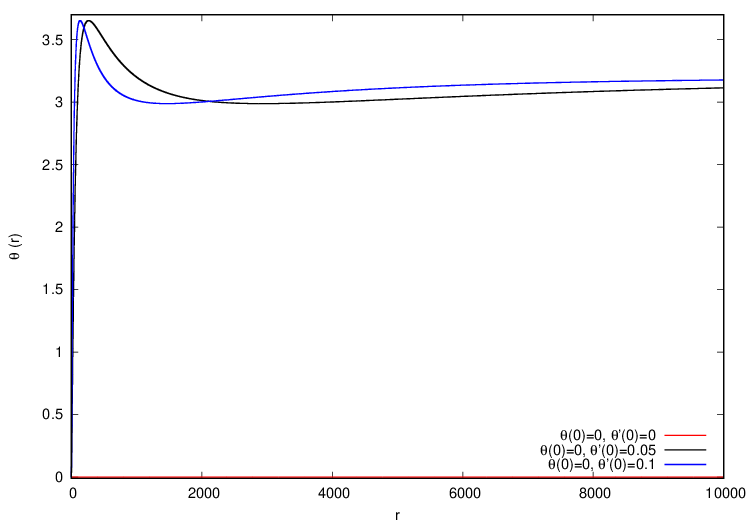}
\vskip12mm
\caption{The numerical solution $\theta$ of the non-linear differential equation \eqref{eq:gfc-perturbative} converging to $\pi$ for the initial conditions at $r=0$: $\theta (0) = 0$ and $\dot{\theta} (0) = 0 \ (\text{red}), \ 0.05 \ (\text{black}), \ 0.1 \ (\text{blue})$.
}
\label{fig:gribov}
\end{figure}

To demonstrate the existence of solutions to the original Gribov equation \eqref{eq:gfc-perturbative}, we have numerically solved it as a non-linear differential equation. The numerical solutions are plotted in Figure~\ref{fig:gribov}.
The numerical solutions for $\theta$ converge to the lowest point, i.e., $\theta = \pi$ as $r \to \infty$ ($\tau \to \infty$) for various initial conditions at $r=0$ ($\tau=-\infty$), since the Gribov equation \eqref{eq:gfc-perturbative} is the equation of motion for the damped pendulum.
Here we have chosen the initial conditions at $r=0$: $\theta (0) = 0$ and $\dot{\theta} (0) = 0 , 0.05 , 0.1$. When the initial velocity is set to $\dot{\theta} (0)=0$, the solution remains at the equilibrium point $\theta = 0$.
These results confirm the above claim based on the solution in the linear approximation.



\section{Restoration of the residual gauge symmetry in the dimensional reduction}

\subsection{Spontaneously breaking of the global $U (1)$ symmetry in the perturbative vacuum}

The perturbative vacuum $\mathscr{A}_\mu = 0$ in the original Yang-Mills theory corresponds to the  reduced field configurations $A_0 = A_1 = \varphi_1 = 0 , \ \varphi_2 = - 1$ according to (\ref{eq:Witten_Ansatz}), namely,
\begin{align}
 \mathscr{A}_\mu(x) = 0 \ (\mu=1,2,3,4) \Leftrightarrow \ A_\mu(r,t) = 0 \ (\mu=0,1), \ \psi(r,t)=\varphi_1(r,t)+i\varphi_2(r,t)=-i=e^{-i \pi/2}.
 \label{eq:pv}
\end{align}
This is indeed the vacuum configuration of the reduced theory (\ref{YM-L}) satisfying $A_\mu(r,t) = 0$ and $|\psi(r,t)|=1$.
Therefore, the global $U(1)$ symmetry $U = e^{-i \theta}$ defined as the global version of  (\ref{eq:gauge_transformation_Witten_Ansatz}) is spontaneously broken by choosing a specific perturbative vacuum (\ref{eq:pv}) with the value $\theta=\pi/2$.
The solution of the Gribov equation (\ref{eq:gfc-perturbative}) also breaks the $U(1)$ symmetry spontaneously in the perturbative vacuum $|0\rangle$:
\begin{align}
 \braket{0|\psi|0} = e^{-i \theta_*},
 \label{eq:pv2}
\end{align}
where $\theta_*$ is the solution of the Gribov equation \eqref{eq:gfc-perturbative} which takes into account the (infinitesimal) global parameter $C_1$ or $C_2$ to define the global $U(1)$ symmetry.
Here it should be remarked that the \textit{local} gauge symmetry cannot be spontaneously broken due to the Elitzur theorem \cite{Elitzur75}.

\subsection{Restoration of the $U (1)$ gauge symmetry and confinement}

\begin{figure}
	\centering
	\includegraphics[width=0.4\textwidth]{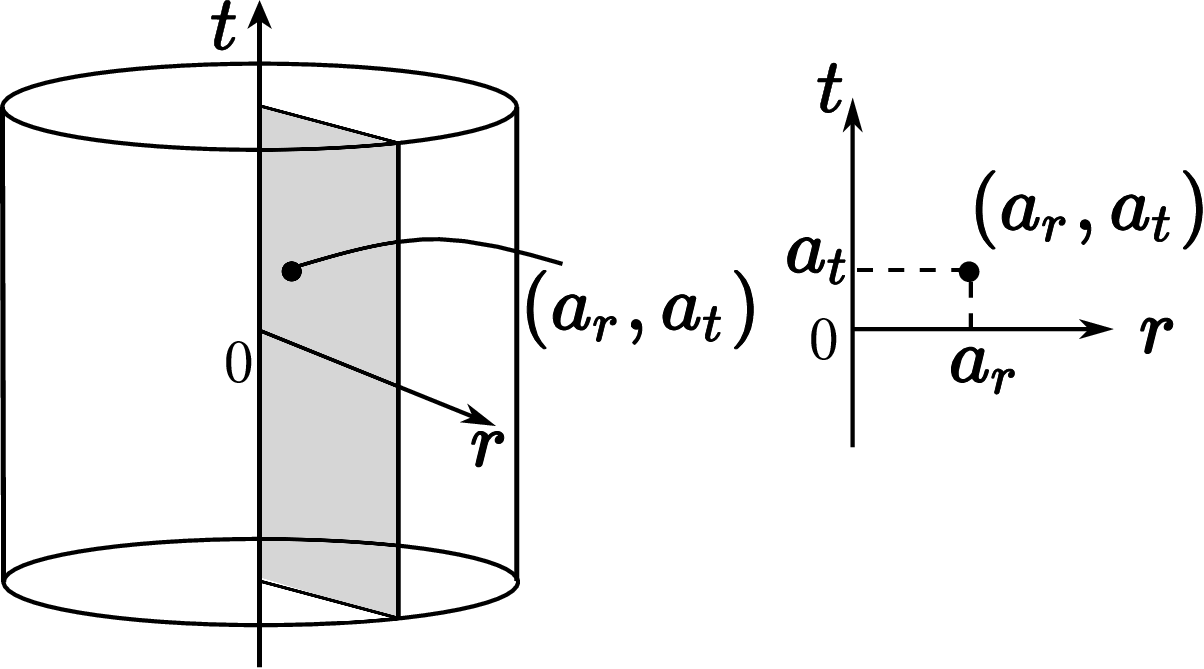}
	\caption{The $(t , r)$ cross section of the cylinder in $\mathbb{R}^4$ and the location $(a_r , a_t)$ of an instanton.}
	\label{fig:cylinder}
\end{figure}

In what follows, we take into account the non-perturbative effects to see the fate of the spontaneously broken global $U(1)$ symmetry indicated by (\ref{eq:pv}) and (\ref{eq:pv2}).
First of all, we start with the equations of motion for the dimensionally reduced $D = 2$ $U (1)$ gauge-scalar theory with the Lagrangian density \eqref{YM-L} given by
\begin{align}
 &- \partial_\mu (r^2 F_{\mu \nu}) + i \psi^* D_\nu \psi - (D_\nu \psi)^* i \psi = 0, \nonumber\\
 &- D_\mu D_\mu \psi + \frac{|\psi|^2 - 1}{r^2} \psi = 0 .
 \label{eq:eom}
\end{align}
Notice that these equations differ from the equations of motion for the  ordinary $U(1)$ gauge-scalar theory by the extra factor $r^2$ which comes from the fact that the relevant $U(1)$ gauge-scalar theory \eqref{YM-L} is defined on the hyperbolic space $\mathbb{H}^2$ specified by the metric $g_{\mu\nu}=r^{-2}\delta_{\mu\nu}$ with a negative constant curvature.
It is quite remarkable to remember that the $U(1)$ gauge-scalar theory \eqref{YM-L} in $D=2$ spacetime has instanton and anti-instanton solutions, see e.g. section 7.14.3 of \cite{Manton}.
However, we do not give the solution explicitly, since we do not need the explicit form of the instanton solution.
In fact, we need just the fact that the 1-instanton solution (with a unit topological charge) approaches asymptotically the vacuum configuration which is obtained through the $U(1)$ gauge transformation from $\psi(r , t)=1$ and $A_\mu(r , t)=0$:
\begin{align}
 &\psi (r , t) = e^{i \theta_a(r , t)} , \quad A_\mu (r , t) = i e^{i \theta_a(r , t)} \partial_\mu e^{- i \theta_a(r , t)} = \varepsilon_{\mu \nu} \frac{(x - a)_\nu}{(x - a)^2} , \nonumber\\
 &\theta_a(r , t) := \arctan \frac{x_1 - a_r}{x_0 - a_t} , \ x=(x_0 , x_1) = (t , r) , \ a=(a_0 , a_1) = (a_t , a_r) ,
 \label{eq:2D_instanton}
\end{align}
where $a=(a_0 , a_1) = (a_t , a_r)$ denotes the location or the center of a 1-instanton on $\mathbb{H}^2$ (see Figure~\ref{fig:cylinder}).
These instanton and anti-instanton solutions are topological soliton solutions characterized by the non-trivial Homotopy group $\pi_1(S^1)=\mathbb{Z}=\{ 0, \pm 1, \pm 2, ... \}$ for the map from the circle $S^1$ winding around the instanton to the target space $U(1) \simeq S^1$.
Then the corresponding topological charge for an $n$-instanton is given by the integer-valued winding number $n$ obtained by
\begin{align}
n := \int d^2x \frac{1}{4 \pi} \varepsilon_{\mu \nu} F_{\mu \nu} \in \mathbb{Z}.
\end{align}

In what follows, we show that the spontaneously broken $U(1)$ symmetry in the perturbative vacuum is restored when the effects of these instantons and anti-instantons are taken into account.
In order to calculate the effects of instantons in general, we need to integrate over all the collective coordinates in the moduli space of instantons, namely, the size and the location of the instanton.
In this case, the size $\lambda$ of an instanton is fixed by the scalar mass $m_s$: $\lambda \approx \mathcal{O} (m_s^{- 1})$ characterized by the scalar potential of the form of the Mexican hat.
Therefore we do not have to integrate over the size of instanton and we have only to consider the integration over the location of instantons.
We adopt the dilute gas approximation for this purpose.
Then the restoration of $U(1)$ symmetry can be demonstrated using the dilute gas approximation of instantons in a way similar to \cite{CDG1977,FMS1980}, as described in detail below.

When the point $x=(r,t)$ of the gauge field $A_\mu (r,t)$ is separated by a distance $d$ larger than the size $\lambda$ from the center $a$ of a 1-instanton,  
the gauge field $A_\mu(r,t)_{a}$ generated by the 1-instanton with the center $a$ has the pure gauge form for $g_{a} \in U(1)$:
\begin{align}
 A_\mu(r,t)_{a} \approx i g_{a} (x) \partial_\mu g_{a}^{- 1} (x), \ g_{a}(x) =e^{i\theta_a(x)} \in U(1).
\end{align}
We consider the configuration of 1-instantons at positions $a_s \ (s = 1 , \cdots n_+)$ and 1-anti-instantons at $b_{s^\prime} \ (s^\prime = 1 , \cdots n_-)$ where $n_+$ and $n_-$ are respectively the total number of 1-instantons and 1-anti-instantons.
We apply the dilute gas approximation where the instantons are assumed to be separated from each other by distances $d$ much larger than their size $\lambda$.
Under this approximation, therefore, the gauge field can be written as the superposition of 1-instantons at $a_s \ (s = 1 , \cdots n_+)$ and 1-anti-instantons at $b_{s^\prime} \ (s = 1 , \cdots n_-)$ (apart from  contributions decaying rapidly in the distance $d$):
\begin{align}
 A_\mu (r,t) \approx \sum_{s = 1}^{n_+} i g_{a_s} (x) \partial_\mu g_{a_s}^{- 1} (x) + \sum_{s^\prime = 1}^{n_-} i g_{b_{s^\prime}} (x) \partial_\mu g_{b_{s^\prime}}^{- 1} (x)
 \quad (|x-a_s|>\lambda, \ |x-b_s|>\lambda).
 \label{2gauge}
\end{align}
In the dilute gas approximation, the action $\tilde S (a_1 , \cdots , a_{n_+} , b_1 , \cdots , b_{n_-})$     and the topological charge $Q(a_1 , \cdots , a_{n_+} , b_1 , \cdots , b_{n_-})$ coming from 1-instantons at $a_s \ (s = 1 , \cdots n_+)$ and 1-anti-instantons at $b_{s^\prime} \ (s = 1 , \cdots n_-)$ can be approximated as
\begin{align}
\tilde S (a_1 , \cdots , a_{n_+} , b_1 , \cdots , b_{n_-}) \approx (n_+ + n_-) \tilde S_0, \quad
Q(a_1 , \cdots , a_{n_+} , b_1 , \cdots , b_{n_-}) = n_+ - n_- ,
\end{align}
where $\tilde S_0$ is the action of the 1-instanton except for a topological term.
Then the $\vartheta$-vacuum is given by
\begin{align}
\ket{\vartheta}
= \sum_{Q = - \infty}^\infty e^{i \vartheta Q} \ket{Q}
= \sum_{n_+ - n_- = - \infty}^\infty e^{i (n_+ - n_-) \vartheta} \ket{n_+ - n_-} ,
\end{align}
where $\ket{n}$ is a vacuum with the winding number $n$.

In this setting, we proceed to calculate the expectation value $\braket{\vartheta| \psi (x)| \vartheta}$ of the scalar field $\psi (x)$ in the $\vartheta$-vacuum $\ket{\vartheta}$ by the path integral formalism.
We start with the theory in a finite spacetime volume $V \subset \mathbb{R}^2$ and finally take the infinite volume limit $V \to \infty$.
First, the partition function including the topological factor $e^{i \vartheta (n_+ - n_-)}$ in a $D = 2$ spacetime volume $V$ is given by
\begin{align}
 Z_V &:= \int_{a_1 , \cdots , a_{n_+} , b_1 , \cdots , b_{n_-} \in V} d A d \psi \ e^{- \tilde S_{GS} [A,\psi]/ \hbar}
 = \sum_{n_+ = 0}^\infty \sum_{n_- = 0}^\infty e^{i \vartheta (n_+ - n_-)}
 \frac{1}{n_+ !} \prod_{s = 1}^{n_+} \int_V \frac{d^2 a_s}{V_0}
 \frac{1}{n_- !} \prod_{s^\prime = 1}^{n_-} \int_V \frac{d^2 b_{s^\prime}}{V_0} \ e^{- (n_+ + n_-) \tilde S_0 / \hbar} \nonumber\\
&= \sum_{n_+ = 0}^\infty \frac{1}{n_+ !} \left( e^{i\vartheta} e^{- \tilde S_0 / \hbar} \frac{V}{V_0} \right)^{n_+} \sum_{n_- = 0}^\infty \frac{1}{n_- !} \left( e^{-i\vartheta} e^{- \tilde S_0 / \hbar} \frac{V}{V_0} \right)^{n_-}
= \exp \left(e^{i \vartheta} e^{- \tilde S_0 / \hbar} \frac{V}{V_0}\right) \exp \left( e^{-i\vartheta} e^{- \tilde S_0 / \hbar} \frac{V}{V_0}\right) \nonumber\\
&= \exp \left(2 \cos \vartheta e^{- \tilde S_0 / \hbar} \frac{V}{V_0}\right),
 \label{eq:partition_function}
\end{align}
where 
$V_0$ is a normalization factor.
In this framework, the $\vartheta$-vacuum expectation value of the scalar field $\psi(x)$ is given by
\begin{align}
 \braket{\vartheta| \psi (x)| \vartheta}_V= &Z_V^{-1} \int_{a_1 , \cdots , a_{n_+} , b_1 , \cdots , b_{n_-} \in V} dA d \psi \ e^{- \tilde S_{GS} [A,\psi]} \psi(x) \nonumber\\
 = &Z_V^{- 1} \sum_{n_+ = 0}^\infty \sum_{n_- = 0}^\infty e^{i \vartheta (n_+ - n_-)} \frac{1}{n_+ !} \prod_{s = 1}^{n_+} \int_V \frac{d^2 a_s}{V_0} \frac{1}{n_- !}  \prod_{s^\prime = 1}^{n_-} \int_V \frac{d^2 b_{s^\prime}}{V_0} \ e^{- (n_+ + n_-) \tilde S_0 / \hbar} [\psi (x)]_{a_1 , \cdots , a_{n_+} , b_1 , \cdots , b_{n_-}} .
\end{align}
where we have defined $[\psi (x)]_{a_1 , \cdots , a_{n_+} , b_1 , \cdots b_{n_-}}$ as the transformed scalar field of a scalar field $\psi_0$ (with $|\psi_0| = 1$ chosen to give a minimum of the scalar potential)
by the same gauge transformations $g_{a_1} , \cdots , g_{a_{n_+}}$ $g_{b_1} , \cdots , g_{b_{n_-}}$ as those used in (\ref{2gauge}):
\begin{align}
 [\psi (x)]_{a_1 , \cdots , a_{n_+} , b_1 , \cdots b_{n_-}} := \prod_{s = 1}^{n_+} g_{a_s} (x) \prod_{s^\prime = 1}^{n_-} g_{b_{s^\prime}} (x) \psi_0.
 \label{eq:psi_transformation}
\end{align}
Then the $\vartheta$-vacuum expectation value of the scalar field $\psi(x)$  is given in the dilute gas approximation by
\begin{align}
 \braket{\vartheta| \psi (x)| \vartheta}_V= &Z_V^{- 1} \sum_{n_+ = 0}^\infty \sum_{n_- = 0}^\infty \frac{e^{i \vartheta (n_+ - n_-)}e^{- (n_+ + n_-) \tilde S_0 / \hbar}}{n_+ ! n_- ! } \prod_{s = 1}^{n_+} \int_{a_s \in V} \frac{d^2 a_s}{V_0} \ g_{a_s} (x) \prod_{s^\prime = 1}^{n_-} \int_{b_{s^\prime} \in V} \frac{d^2 b_{s^\prime}}{V_0} \ g_{b_{s^\prime}} (x) |\psi_0|  .
\end{align}

\begin{figure}
	\centering
	\includegraphics[width=0.8\textwidth]{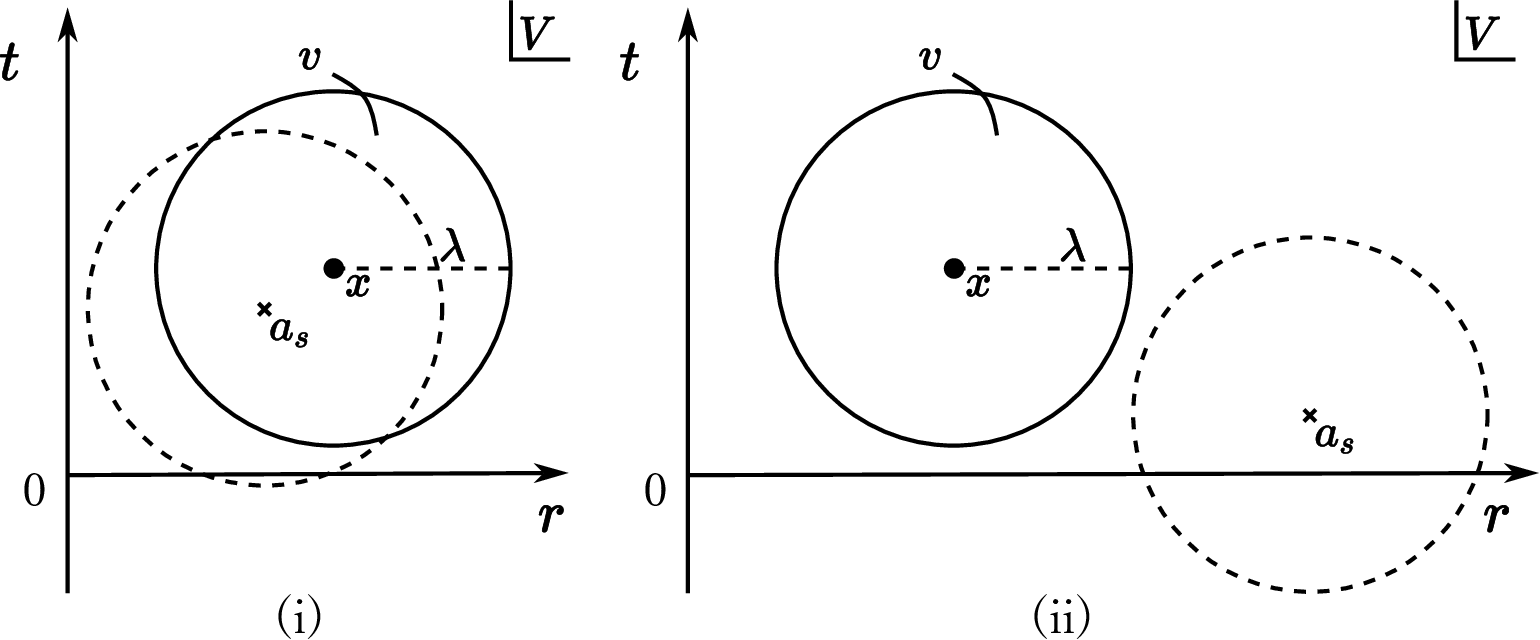}
	\caption{
	The decomposition of the integration region with a given (fixed) $x$ into two parts in \eqref{eq:Vv}.
	({\rm i}) The center $a_s$ of a 1-instanton is inside the region $v$ centered at $x$; $|a_s - x| < \lambda$:
	The instanton centered at $a_s$  includes the point $x$ in the inside.
	({\rm ii}) The center $a_s$ of a 1-instanton is outside the region $v$ centered at $x$; $|a_s - x| > \lambda$:
	The instanton centered at $a_s$ does not include the point $x$  in the inside.
}
	\label{fig:Vv}
\end{figure}

Next, we proceed to estimate the integration over the locations of instantons.
Let $v = \mathcal{O} (\lambda^2)$ be the volume, i.e., the area satisfying $\{ a_s \in V; |a_s - x| < \lambda \}$ and $\{ b_{s^\prime} \in V; |b_{s^\prime}- x| < \lambda \}$ for a given $x\in V$.
We find there exist some $\eta$ and  $\xi$ satisfying $0 < \xi < 1$ and $0 <\eta<1$ such that the following bounds hold uniformly in $V$ (for $V$ large enough): 
\begin{align}
 \left|\int_{a_s \in V , \ |a_s - x| < \lambda} d^2 a_s \ g_{a_s} (x)\right| \leq v \eta , \quad
 \left|\int_{a_s \in V , \ |a_s - x| > \lambda} d^2 a_s \ g_{a_s} (x)\right| \leq (V - v) \xi .
\end{align}
These bounds hold also by replacing $a_s$ with $b_{s^\prime}$.
  In order to estimate the integral over the location of instantons in the whole (finite) space $V$, we decompose the integral over the whole space $V$ into the two parts (see Figure~\ref{fig:Vv}.): ({\rm i}) $a_s$ is inside the region $v$ centered at $x$ and the instanton with the center $a_s$ overlaps with $x$, ({\rm ii}) $a_s$ is outside the region $v$ centered at $x$ and the instanton with the center $a_s$ does not overlap with $x$:
\begin{align}
 \left|\int_{a_s \in V} d^2 a_s \ g_{a_s } (x)\right| = & \Biggl|\int_{a_s \in V , \ |a_s - x| > \lambda} d^2 a_s \ g_{a_s } (x) + \int_{a_s \in V , \ |a_s - x| < \lambda} d^2 a_s \ g_{a_s } (x)\Biggr| \nonumber\\
 \leq & \left|\int_{a_s \in V , \ |a_s - x| > \lambda} d^2 a_s \ g_{a_s } (x)\right| 
 + \left|\int_{a_s \in V , \ |a_s - x| < \lambda} d^2 a_s \ g_{a_s } (x)\right| 
 \leq  \left(V - v\right) \xi + v \eta .
 \label{eq:Vv}
\end{align}
The same bound is obtained for anti-instantons by replacing $a_s$ with $b_{s^\prime}$.
Therefore, we estimate the upper bound on $|\braket{\vartheta| \psi (x)| \vartheta}_V|$:
\begin{align}
  |\braket{\vartheta| \psi (x)| \vartheta}_V|
  \leq & Z_V^{- 1} \sum_{n_+ = 0}^\infty  \sum_{n_- = 0}^\infty \frac{1}{n_+ ! n_- !} |e^{i \vartheta (n_+ - n_-)}| e^{- (n_+ + n_-) \tilde S_0 / \hbar} \left(\frac{V}{V_0}\right)^{n_+ + n_-} \left[\left(1 - \frac{v}{V}\right) \xi + \frac{v}{V} \eta\right]^{n_+ + n_-} |\psi_0|  \nonumber\\
=& \exp \left(-2 \cos \vartheta e^{- \tilde S_0 / \hbar} \frac{V}{V_0}\right)
\exp \left(2 e^{- \tilde S_0 / \hbar} \frac{(V-v) \xi + v\eta}{V_0}  \right) |\psi_0| \nonumber\\
 = & \exp \left(- 2 e^{- \tilde S_0 / \hbar} \frac{V}{V_0} (\cos \vartheta - \xi)\right) \exp \left(2 e^{- \tilde S_0 / \hbar} \frac{v}{V_0} (- \xi + \eta)\right) |\psi_0| .
\end{align}
This upper bound decreases to zero as $V \to \infty$ as long as the $\vartheta$ angle is small $|\vartheta| \ll 1$ such that $\cos \vartheta>\xi$ for any $\xi<1$, which gives $2 e^{- \tilde S_0 / \hbar} \frac{V}{V_0} (\cos \vartheta - \xi) > 0$.
The value of $\vartheta$ in the $D=4$ Yang-Mills theory is usually restricted to be very small not to cause the violation of the strong CP  invariance, although the above argument does not exclude the possibility of non-restoration of RGS for large value of $\vartheta$ from the purely theoretical side.
The contribution $v \eta$ from inside the region $v$ does not affect the limit, regardless of the value of $\eta$.
Thus we have demonstrated the restoration of the $U (1)$ symmetry:
\begin{align}
 \braket{\vartheta| \psi (r , t)| \vartheta} = \lim_{V \to \infty}\braket{\vartheta| \psi (r , t)| \vartheta}_V = 0 .
 \label{eq:restoration0}
\end{align}
In this case, the RGS for $U = \exp \left(i \theta \frac{x_A}{r} \frac{\sigma_A}{2}\right)$ must satisfy the Gribov equation:
\begin{align}
  \partial_\mu (r^2 \partial_\mu \theta (r,t)) = 0.
\end{align}

Now we consider the $\vartheta$-vacuum expectation value $\braket{\vartheta| A_\mu (x)| \vartheta}$ of the gauge field $A_\mu$.
When the point $x=(r,t)$ of the gauge field $A_\mu (r,t)$ is separated by a distance larger than the size $\lambda$ from the center $a$ of a 1-instanton and the center $b$ of a 1-anti-instanton,
the gauge field $A_\mu(x)$ can be expressed as a sum of 1-instantons and 1-anti-instantons (\ref{2gauge}) in the dilute gas approximation.
In this case, there always occurs the cancellation between the contribution from an instanton located at $a$ and an anti-instanton at $b$ and the contribution from the exchanged configuration, i.e., an anti-instanton at $a$ and an instanton at $b$.
In the other case, the integral $\int_{a \in V, \ |a - x| < \lambda} d^2 a \ A_{\mu}(x)_a$ is performed over the region $v$.
Since this is localized in the region $v$, the whole volume $V$ as the integration domain is irrelevant for the result of integration and the limit $V \to \infty$ does not affect the integration result. Consequently, the $\vartheta$-vacuum expectation value of the gauge field $A_\mu$ vanishes in the similar way to $\braket{\vartheta | \psi(x) | \vartheta}$:
\begin{align}
  \braket{\vartheta| A_\mu(r , t) | \vartheta} = \lim_{V \to \infty} \braket{\vartheta | A_\mu(r , t) | \vartheta}_V = 0 .
  \label{eq:restoration1}
\end{align}

It should be noticed that the recovery of the global $U(1)$ symmetry we have just shown is consistent with the Mermin-Wagner-Coleman theorem \cite{MW66,Coleman73}: The absence of spontaneous breaking of continuous global symmetries in $D \le 2$ dimensional infinite systems, which is different from the Elitzur theorem for the local symmetry which holds irrespective of the dimensionality $D$.

\section{Conclusion and discussion}

In this paper, we have demonstrated that the restoration of the RGS in the Lorenz gauge for the $D=4$ $SU(2)$ Yang-Mills theory indeed occurs at least if the theory is restricted to the sector respecting the spatial spherical symmetry $SO(3)$, which causes the dimensional reduction to the $D = 2$ $U(1)$ gauge-scalar theory.

By this reduction, the spontaneous breaking of the RGS and its restoration in the $D=4$ $SU(2)$ Yang-Mills theory is reduced to the spontaneous breaking of the global $U(1)$ symmetry and its restoration in the reduced $D = 2$ $U(1)$ gauge-scalar theory.
The restoration is realized by non-perturbative effects due to instantons (topological defects usually called vortices) in the reduced $D = 2$ $U(1)$ gauge-scalar theory which are different from the $D = 4$ Yang-Mills instantons with spacetime spherical symmetry.
Therefore, our result is consistent with the conventional understanding that the $D = 4$ Yang-Mills instantons do not directly contribute to confinement in the $D=4$ Yang-Mills theory.

Moreover, it is known \cite{CDG1977} that the instantons in $D = 2$ $U(1$) gauge-scalar theory lead to quark confinement in the sense of Wilson \cite{Wilson74}, i.e., area law of the Wilson loop average which means the linear potential for the static quark potential.

Indeed, under the Witten Ansatz, the Wilson loop operator in any representation in $D = 4$ Yang-Mills theory is reduced to the corresponding Abelian Wilson-like operator in $D = 2$ Abelian gauge-scalar theory by using the non-Abelian Stokes theorem. The exact relationship is explained together with the derivation of the area law in \cite{Kondo25}. Therefore, investigating such an operator in the dimensionally reduced $D = 2$ Abelian gauge-scalar theory enables us to revisit and investigate the Wilson loop in the $D = 4$ non-Abelian gauge theory.


It should be remarked that even in the Abelian gauge theory confinement occurs.
For any spacetime dimensions $D \ge 2$, the Abelian and non-Abelian gauge theories  with \textit{compact}  gauge groups including the Abelian group $U(1)$ exhibit confinement at least in the strong coupling region, as was shown in the famous paper by K.G. Wilson  \cite{Wilson74}. Here confinement is demonstrated by  evaluating the vacuum expectation value (average) of the Wilson loop operator to derive the area law of the Wilson loop average, i.e., the existence of the linear potential $V(r) =\sigma r$ in the quark-antiquark potential using the strong coupling expansion in the strong coupling region in the framework of the lattice gauge theory. The strong coupling expansion on the lattice is a convergent infinite series with a finite radius of convergence and hence the result is mathematically rigorous in sharp contrast to the perturbative expansion which is a divergent series and mathematically ill-defined.

For $D=3$, even the ordinary \textit{non-compact} $U(1)$ gauge theory, i.e. continuum quantum Maxwell theory exhibits confinement for any value of the coupling constant with the logarithmic potential $V(r) \propto \log r \to \infty (r \to \infty)$ instead of the linear potential.
For $D=4$, the Abelian gauge theory with \textit{compact} $U(1)$ gauge group has the confinement phase in the strong coupling region which is separated by a critical coupling from the Coulomb phase in the weak coupling region in the framework of the lattice gauge theory, as shown e.g., by
\cite{Polyakov77,BMK77}.
Here, it is demonstrated that magnetic monopoles exist in the strong coupling region and that the condensation of magnetic monopoles causes the dual Meissner effect leading to confinement.
These results show that topological effects such as monopole condensation realized through the compactification of the internal space, leads to area law of the Wilson loop average corresponding to the linear potential between charges.

It should be noted that the linear potential in $D=2$ is a quite non-trivial result due to quantum effects which vanish in the classical limit $\hbar \to 0$ and does not follow from the trivial fact that the flux of gauge field is restricted to one spatial dimension for $D=2$ which is nothing but the classical aspect.
This fact supports the expected intimate relationship between confinement and the restoration of RGS.

We give a comment on the sign of the beta function in the relationship with confinement.
It is known that the beta function for a class of quantum field theories can be negative in the spacetime dimension $D$ less than four $D < 4$, even if it is positive in $D = 4$, as
 clearly explained in Section 12.5 of \cite{Peskin}.
For example, the beta function $\beta(\lambda)$ of the scalar $\lambda \varphi^4$ theory with the coupling constant $\lambda$ in the spacetime dimension $D$ is given by
\begin{align}
  \beta(\lambda) = - (4 - D) \lambda + \frac{3 \lambda^2}{16 \pi^2}.
\end{align}
Therefore, for $D=4$  the beta function $\beta(\lambda)$ is positive $\beta(\lambda)=\frac{3 \lambda^2}{16 \pi^2}>0$.
The term $- (4 - D) \lambda$ appears only for $D \neq 4$ and is negative for $D < 4$ in the region $\lambda>0$.
Therefore, the beta function becomes negative $\beta(\lambda) < 0$ which is reliable in the region of the sufficiently small $\lambda$, apart from the possible existence of the infrared-stable fixed point for a certain finite coupling $\lambda_*$ (see Fig.~12.4 of \cite{Peskin}).

The appearance of such a negative term in the beta function for $D < 4$ is not a special feature of the scalar $\lambda \varphi^4$ theory, but it is common to the other quantum field theories which are renormalizable in $D=4$ where the coupling constant is dimensionless in $D=4$, while it has the non-vanishing dimension for $D < 4$ (e.g., $4-D$ for $\lambda$ of $\lambda \varphi^4$ theory).
In fact, the beta function of the Abelian gauge theory, e.g., $U(1)$ gauge-scalar theory with the gauge coupling constant $e$ exhibits the similar behavior:
\begin{align}
  \beta(e) = - \frac{4-D}{2} e + \frac{e^3}{48 \pi^2}.
\end{align}
Thus, the negativity of the beta function for $D < 4$ is consistent with the occurrence of confinement even in the Abelian gauge theory for $D < 4$, which is comparable with the expected confinement in the $D=4$ renormalizable non-Abelian gauge theory with the negative beta function showing ultraviolet asymptotic freedom.

We are interested in extending the analysis of the RGS as a confinement criterion to the other gauges.
In this paper, the restoration of the RGS in the $SU(2)$ Yang-Mills theory in the Lorenz gauge was shown by focusing on the $U(1)$ subgroup of the original $SU(2)$ gauge group which is obtained through the dimensional reduction caused by restricting the Yang-Mills field to symmetric instantons.
Therefore, if we recall that the Maximal Abelian gauge restricts the original non-Abelian gauge group to its maximal torus subgroup, e.g. $SU(2)$ to $U(1)$, our framework given in this paper will be able to provide a suitable framework to study the RGS also in this gauge as initiated in \cite{Kondo-Fukushima}.

Finally, we give a comment on the Kugo-Ojima criterion for color confinement in the Lorenz gauge which tells us that the specific value $u(k=0)=-1$ of the Kugo-Ojima function $u(k)$ in the infrared limit is sufficient to conclude confinement of all colored objects.
However, the Kugo-Ojima function is identically zero for the Abelian gauge group, which means that the Abelian gauge theory does not exhibit confinement.
According to Kugo-Ojima, therefore, the dimensionally reduced U(1) gauge-scalar theory does not show confinement.
It should be remembered that the Kugo-Ojima criterion was derived from the requirement of the well-definedness of the color charge operator which follows from the absence of the massless pole contributions.
It is well known that massless particles cannot exist in the two-dimensional spacetime.
Therefore, we must understand that the Kugo-Ojima criterion cannot be applied to the two-dimensional case, namely, the reduced $D=2$ theory.
Therefore, it is not strange that the restoration of the RGS and the Kugo-Ojima criterion give different results.
This result suggests that two criteria are in general different as confinement criterion.

In this paper, we focused on the case of the $SU (2)$ gauge group. In order to discuss the real QCD, therefore we should extend our scheme to the $SU (3)$ gauge theory. The Witten Ansatz corresponds to the special case where the color direction field $n^A = \frac{x_A}{r}$ is taken in the field decomposition method proposed by Faddeev and Niemi \cite{FN_to_WA,FN}, as mentioned in \cite{FN_to_WA}. This field decomposition method has been proposed not only for $SU (2)$ but also for general $SU (N)$.
The possible choice of the color direction field for $SU (3)$ has been exhausted in \cite{PR}.
Therefore, it is in principle applicable to $SU (3)$. However, since the relationship between the choice of the color direction field and the dimensional reduction is still to be investigated in the $SU (3)$ case, the details in real QCD will be left as a future work.

\section*{Acknowledgements}
This work was supported by Grant-in-Aid for Scientific Research, JSPS KAKENHI Grant Number (C) No.23K03406.
N.F. was supported by JST SPRING, Grant Number JPMJSP2109.


\bibliographystyle{elsarticle-harv}


\appendix
\section{Implications of the $D=2$ symmetry restoration for the original $D = 4$ $SU (2)$ Yang-Mills theory}

In this Appendix, we discuss the implications of the $D=2$ statements \eqref{eq:restoration0} and \eqref{eq:restoration1} for the original $D = 4$ $SU (2)$ Yang-Mills theory.
By taking the $\vartheta$-vacuum expectation value of both sides of the Witten transformation (\ref{eq:Witten_Ansatz}), we find that the restoration of RGS \eqref{eq:restoration0} and \eqref{eq:restoration1} implies
\begin{align}
 \braket{\vartheta| \mathscr{A}_4^A (x) | \vartheta} =& -\frac{x_A}{r} \braket{\vartheta| A_0 (r , t) | \vartheta}  = 0 , \nonumber\\
 \ \braket{\vartheta| \mathscr{A}_j^A (x) | \vartheta} =&- \frac{x_j x_A}{r^2} \braket{\vartheta| A_1 (r , t) | \vartheta}  - \frac{\delta_{j A} r^2 - x_j x_A}{r^3} \braket{\vartheta| \varphi_1 (r , t) | \vartheta}  - \frac{\varepsilon_{j A k} x_k}{r^2} [1 + \braket{\vartheta| \varphi_2 (r , t) | \vartheta} ]
 = \frac{\varepsilon_{A j  k} x_k}{r^2} \quad (j = 1,2,3).
 \label{eq:Witten_Ansatz2}
\end{align}

Then the $\vartheta$-vacuum expectation value of the original $SU(2)$ Yang-Mills field resulting from the restoration of RGS \eqref{eq:restoration0} and \eqref{eq:restoration1} agrees with the profile function of the static magnetic monopole called the Wu-Yang magnetic monopole \cite{WY-monopole}:
\begin{align}
 \mathscr{A}_4^A (x)=0, \quad \mathscr{A}_j^A (x)=\varepsilon_{Ajk} \frac{x_k}{r^2} .
\end{align}
This is a special case $f(r)\equiv 0$ of the  't Hooft-Polyakov magnetic monopole \cite{tP-monopole} and the magnetic monopole of the massive Yang-Mills theory \cite{Nishino2018} with the profile function $f(r)$ with the boundary condition $f(r\to 0) \to 1$ and the asymptotic behavior  $f(r\to \infty) \to e^{-Mr} \sim 0$ for the Ansatz for the gauge field $\mathscr{A}_\mu^A (x)$ and the scalar field $\phi^A(x)$:
\begin{align}
 \mathscr{A}_4^A (x)=0, \quad \mathscr{A}_j^A (x)=\varepsilon_{Ajk} \frac{x_k}{r^2} [1-f(r)], \quad \phi^A(x) = \frac{x^A}{r} h(x) ,
\end{align}
It is usually understood that the 't Hooft-Polyakov monopole is formed as a result of spontaneous symmetry breaking where the Yang-Mills field for $SU(2)/U(1)$ part acquire the mass $M=gv$ (the Higgs mechanism) by absorbing the massless Nambu-Goldstone particle associated with the partial spontaneous symmetry breaking $SU(2) \to U(1)$ as a consequence of the nonvanishing vacuum expectation value $\langle \phi^A(x)\rangle=\frac{x^A}{r}v$ of the scalar field, while the $U(1)$ part remains massless since $U(1)$ symmetry remains unbroken.
Consequently, there occurs the length scale $O(M^{-1})$ coming from the generated mass $M$ of the gauge field so that $f(r) >0 \ (r <M^{-1})$ and $f(r) \cong 0\  (r > M^{-1})$, see e.g., \cite{Nishino2018}.
The $SU(2)$ symmetry is restored at the center of the magnetic monopole where the scalar field vanishes, while in the long distance $r>O(M^{-1})$ $SU(2)$ is broken to $U(1)$ with very small value of the scalar field.
The above result indicates that the 't Hooft-Polyakov magnetic monopole or the Yang-Mills magnetic monopole is converted to the Wu-Yang monopole in the whole region as the result of the dimensional reduction.

The reduction (\ref{eq:gauge_transformation_Witten_Ansatz}) of the gauge group from $U(x) = \exp \left(i \theta(x) \text{n}^A(x) \frac{\sigma_A}{2}\right) \in SU(2)$ to $U(x) =\exp (i\theta(x) )$ is unavoidable as long as the unit vector field $n^A(x)$ is chosen to be a fixed vector $n^A(x) = \frac{x_A}{r}$ (\ref{eq:choiceofn}) in agreement with the Witten transformation.
In order to fully recover the original symmetry, we need to take into account the full degrees of freedom of $n^A(x)$ as a field variable without restricting to a fixed vector such that the field $n^A(x)$ transforms according to the adjoint representation of the gauge group, see \cite{PR} for details.
Up to this restriction, thus, our result is consistent with the restoration of the RGS in the $D = 4$ Yang-Mills theory.


\end{document}